\documentclass[12pt]{article}
\usepackage{epsfig}

\def\beq{\begin{equation}}
\def\eeq#1{\label{#1}\end{equation}}
\def\eeqn{\end{equation}}

\def\beqa{\begin{eqnarray}}
\def\eeqa#1{\label{#1}\end{eqnarray}}
\def\eeqan{\end{eqnarray}}

\let\bar=\overbar

\def\Dslash{\not{\hbox{\kern-4pt $D$}}}
\def\dslash{\not{\hbox{\kern-2pt $\del$}}}

\def\msb{{\bar{\ssstyle M \kern -1pt S}}}

\def\Title#1{\begin{center} {\Large {\bf #1} } \end{center}}

\begin{document}

\Title{$R_K$ and $K^+\rightarrow\pi^+\nu\bar{\nu}$ with NA62 at CERN SPS\footnote{Proceedings of CKM 2012, the 7th International Workshop on the CKM Unitarity Triangle, University of Cincinnati, USA, 28 September - 2 October 2012.}}

\bigskip\bigskip

\begin{raggedright}  

{\it Giuseppe Ruggiero\index{Ruggiero, G.}\footnote{On behalf of NA62 collaboration.}\\
CERN, PH\\
Geneva, Switzerland}
\bigskip\bigskip
\end{raggedright}

\section{Introduction}
In the Standard Model (SM) the decays of charged pseudo scalar mesons into lepton neutrino ($P^\pm\rightarrow l^\pm\nu$, denoted $P_{l2}$) are helicity suppressed. 
The SM allows a very precise determination of the ratio of $P_{l2}$ decay amplitudes of the same meson in different lepton flavours, like $R_K\equiv\Gamma(K_{e2})/\Gamma(K_{\mu2})$.
The prediction for $R_K$ inclusive of internal bremsstrahlung (IB) radiation is \cite{Cirigliano:2007xi} $R_K=(2.477\pm0.001)\times10^{-5}$.
%%\begin{equation}
%%R_K^{SM}\,=\,\left(\frac{M_e}{M_\mu}\right)^2\left(\frac{M^2_K-M^2_e}{M^2_K-M^2_\mu}\right)^2(1+\delta R_{QED})\,=\,(2.477\pm0.001)\times10^{-2}, 
%%\end{equation}
%%where $\delta R_{QED}$ is an electromagnetic correction.
Deviations of $R_K$ from the SM require new physics models with sources of lepton flavour violation (LFV) \cite{Masiero:2005wr,Masiero:2008cb,Recent}. 
%A recent study \cite{recent}
%showed that $R_K$ can be enhanced up to \% level within the Minimal Supersymmetric Standard Model. Other observables like $B_s\rightarrow\mu^+\mu^-$ and $B_d\rightarrow\tau\nu$,
%however, may constraint the potential new physics effects on $R_K$ \cite{others}.
The present PDG value \cite{pdg} of $R_K$, $R_K=(2.488\pm0.012)\times10^{-5}$, is largely dominated by the result 
%from KLOE \cite{Ambrosino:2009rv} and 
of the analysis of a partial sample of the data set collected by the NA62 experiment at CERN in 2007 \cite{Lazzeroni:2011}. 
A new measurement of $R_K$ based on the full data sample collected by NA62 is reported here (section \ref{sec:rk}).

Among the many rare flavour changing neutral current $K$ and $B$ decays, the ultra rare decays $K\rightarrow\pi\nu\bar{\nu}$ play a key role in the search for new physics through underlying 
mechanisms of flavour mixing. The SM branching ratio can be computed to an exceptionally high degree of precision.
%: the $O(G^2_F)$ electroweak amplitudes exhibit a power-like GIM mechanism; 
%the top-quark loops largely dominate the matrix element; the sub-leading charm-quark contributions have been computed at NNLO order \cite{Buras:2006gb}; the hadronic matrix element can be 
%extracted from the branching ratio of the $K^+\rightarrow\pi^0e^+\nu$ decay, well known experimentally \cite{Ams:2008}. 
The prediction for the $K^+\rightarrow\pi^+\nu\bar{\nu}$ channel is $(7.81\pm0.75\pm0.29)\times10^{-11}$ \cite{Brod:2010hi}. The first error comes from the uncertainty on the CKM matrix elements, 
the second one is the pure theoretical uncertainty. This decay is one of the best probe for new physics effects complementary to the LHC, expecially within non Minimal Flavour 
Violation models \cite{Isidori:2006qy,Blanke:2009am}. Since the extreme theoretical cleanness of these decays remains also in these scenarios, even deviations from the SM value at the level of 20\% 
can be considered signals of new physics. Also, the decay can be used for a measurement of $V_{td}$ free from hadronic uncertainties and independently from that obtained with $B$ mesons decays. The decay 
$K^+\rightarrow\pi^+\nu\bar{\nu}$ has been observed by the experiments E787 and E949 at the Brookhaven National Laboratory and the measured branching ratio is $1.73^{+1.15}_{-1.05}\times10^{-10}$ 
\cite{Artamonov:2009sz}. However only a measurement of the branching ratio with at least 10\% accuracy can be a significant test of new physics. This is the main goal of the NA62 experiment at CERN-SPS 
\cite{na62tdr} (section \ref{sec:pvv}). 

\section{Measurement of $R_K$ with NA62}
\label{sec:rk}
The NA62 experiment at CERN collected data in 2007-08 for measuring $R_K$ using the same beam line and experimental set-up of the NA48/2 experiment \cite{Batley:2007yfa}.
The experiment made use of a 400 GeV/c primary proton beam, extracted from the SPS accelerator at CERN and producing a secondary charged kaon beam after impinging on 
a beryllium target. A 100 m long beam line selected the momentum of the secondary beam to $(75\pm3)$ GeV/c. Finally the beam entered a decay volume, housed in a 100 m long vacuum tank. 
The detector was designed to see the products of the kaons decaying in the vacuum region. A magnetic spectrometer, consisted of four drift 
chambers separated by a dipole magnet, tracked the charged particles. 
% It provided a transverse momentum kick of 265 MeV/c, corresponding to a momentum resolution of $\sigma_p/p=(0.48\oplus0.009\times p)$\% ($p$ in GeV/c).
An array of scintillators (hodoscope) gave the time reference for the other detectors and the main trigger for the event topologies including charged particles.
%An hodoscope, made of two orthogonal planes of 64 plastic scintillator slabs each, gave the time reference for the other detectors and the main trigger 
%for the events with charged particles. 
%An electromagnetic calorimeter (LKr), placed after the hodoscope, was used for photon detection and particle identification. It is a quasi-homogeneous calorimeter 
%with liquid kripton as active material. The measured energy resolution is $\sigma(E)/E=0.032/\sqrt(E)\oplus0.09/E\oplus0.0042$ ($E$ in GeV). 
A quasi-homogeneous electromagnetic calorimeter with liquid Kripton as active material (LKr) \cite{Fanti}, was used mainly for particle identification. 
%It is a quasi-homogeneous calorimeter 
%with liquid kripton as active material. The measured energy resolution is $\sigma(E)/E=0.032/\sqrt(E)\oplus0.09/E\oplus0.0042$ ($E$ in GeV). 

The analysis strategy for measuring $R_K$ is based on counting the numbers of reconstructed $K_{e2}$ and $K_{\mu2}$ events collected concurrently. Consequently $R_K$ does not depend on the absolute kaon 
flux and the ratio allows for a first order cancellation of several systematic effects, like reconstruction and trigger efficiencies and time dependent biases. The basic formula is:
\begin{equation}
R_K\,=\,\frac{1}{D}\cdot\frac{N(K_{e2})-N_B(K_{e2})}{N(K_{\mu2})-N_B(K_{\mu2})}\cdot\frac{A(K_{\mu2})f_\mu\epsilon(K_{\mu2})}{A(K_{e2})f_e\epsilon(K_{e2})}\cdot\frac{1}{f_{LKr}}.
\end{equation}
Here $N(K_{l2})$ and $N_B(K_{l2})$ are the number of selected $K_{l2}$ events and expected background events, respectively; $D$ is the downscaling factor applied to the $K_{\mu2}$ trigger; 
$A(K_{l2})$ the geometrical acceptance of the selected $K_{l2}$ mode; $f_e$ and $f_\mu$ the identification efficiencies of electrons and muons, respectively; $\epsilon(K_{l2})$ the trigger efficiencies for the 
selected $K_{l2}$ events; $f_{LKr}$ the global LKr efficiency. A Monte Carlo simulation is used to evaluated the acceptance correction and the geometric part of the acceptances for most of the background processes. 
The beam halo background, the particle identification and the readout and trigger efficiencies are measured directly from data. Both $f_{l}$ and $\epsilon(K_{l2})$ are well above 99\%.  
The analysis is performed independently for 40 data samples (10 bins of reconstructed lepton momentum and 4 samples with different data taking conditions).

A large part of the selection is in common between $K_{e2}$ and $K_{\mu2}$, because of the similar single-track topology. The event kinematics and the lepton identification, instead, are effective to
separate $K_{e2}$ and $K_{\mu2}$. The kinematic identification is based on the squared missing mass $M^2_{miss}=(P_{K}-P_{l})^2$, where $P_K$ and $P_l$ are the kaon and lepton 4-momenta respectively.
The mean $P_K$ is monitored on spill basis using fully reconstructed $K^+\rightarrow\pi^+\pi^+\pi^-$decays; $P_l$ is computed in the electron or muon mass hypothesis. A cut on the $M^2_{miss}$, according to the 
$M^2_{miss}$ resolution and dependent on the lepton momentum, selects the $K_{l2}$ candidates.

\begin{figure}[htb]
\centering
\includegraphics[width=0.4\textwidth]{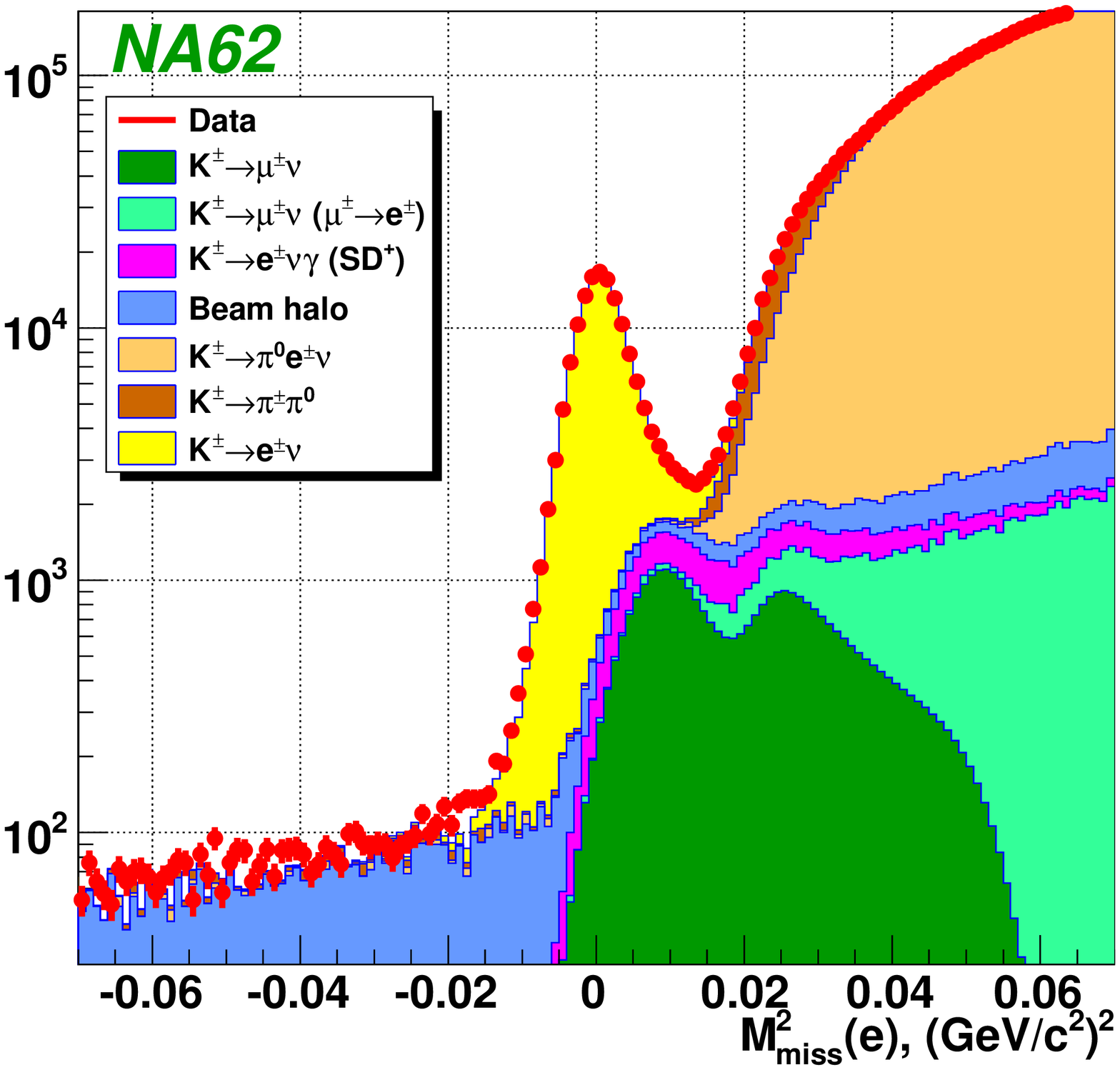}
\includegraphics[width=0.4\textwidth]{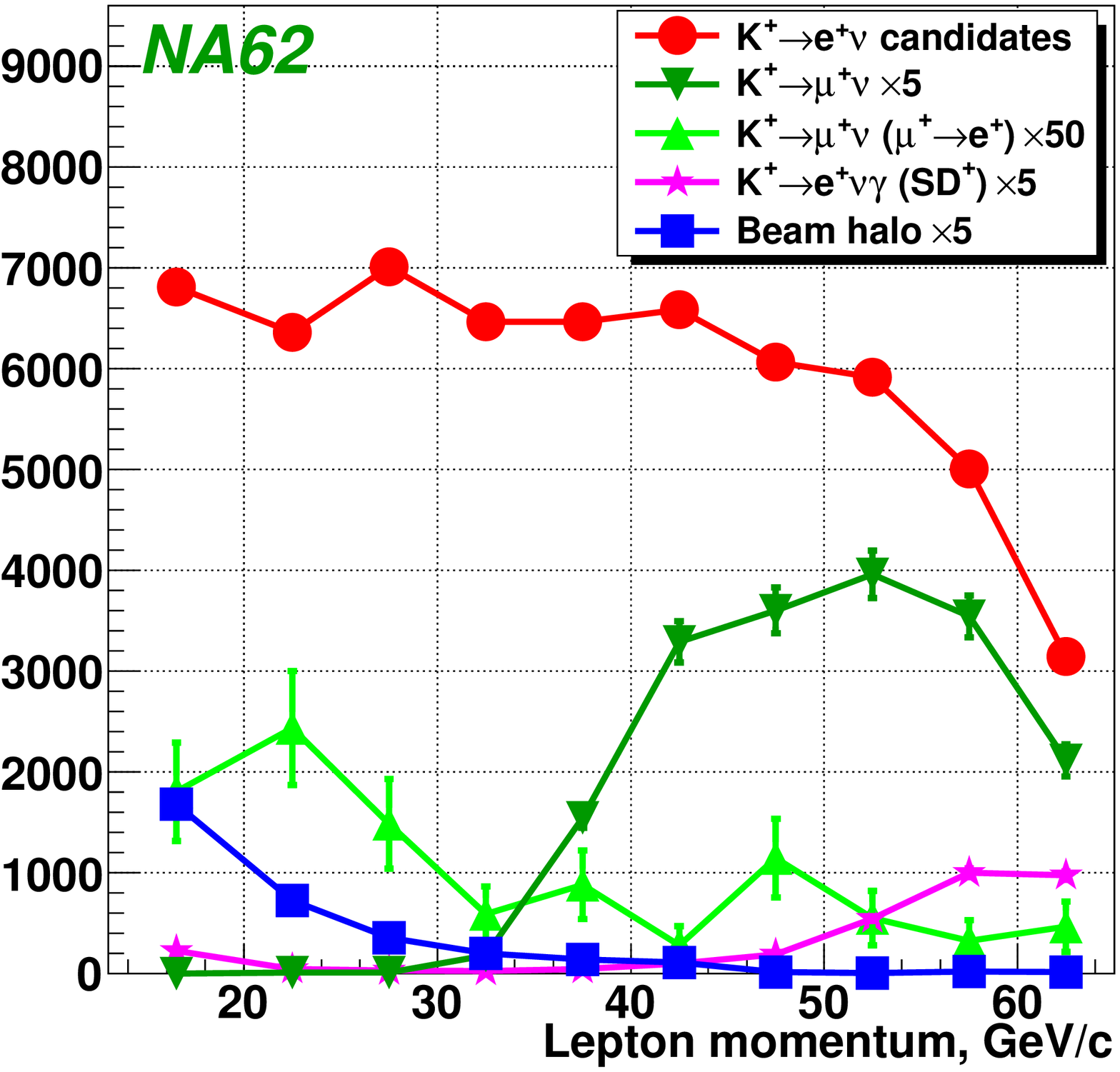}
\caption{Left: reconstructed $M^2_{miss}(K_{e2})$ for $K_{e2}$ and backgrounds. Right: lepton momentum distributions of the $K_{e2}$ candidates and the dominant
backgrounds.}
\label{fig:m2miss}
\end{figure}

The numbers of selected $K_{e2}$ and $K_{\mu2}$ candidates are $145958$ and $4.2817\times10^7$, respectively. The total background is $10.95\pm0.27$\%. The $M^2_{miss}(K_{e2})$ and the background contamination as a 
function of lepton momentum are shown in figure \ref{fig:m2miss}. 

The background is strongly momentum dependent. In particular for momenta higher than 35-40 GeV/c, the $K_{\mu2}$ with a muon mis-identified as a positron becomes the largest background source. 
%The accuracy of its evaluation is critical to keep the total systematic uncertainty smaller than the statistical one. 
A large energy loss as a consequence of a hard bremsstrahlung of the muon in or in front of the LKr is the dominant source of the muon-positron mis-identification. The corresponding probability 
is measured on data as a function of lepton momentum by making use of a set of data collected in 2007 with a 9.2 $X_0$ lead bar in front of the LKr, covering about 20\% of the total geometrical 
acceptance. This set-up allowed the collection of a muon sample free from electron contamination due to $\mu\rightarrow e$ decays. 
%The measured probability is in the range of $10^{-6}$ according to the muon momentum and is in agreement with the simulation within the uncertainties (about 10\% from simulation). 
%The other background components coming from kaon decays were evaluated using the MC simulation. 
%The beam halo background induced by halo muons undergoing decay in flight or mis-identified, was measured by reconstructing positive $K_{e2}$ among data collected from 
%$K^-$ beam with the $K^+$ beam blocked while its halo was not. The size of the control sample limited the evaluation of the background uncertainty. 
%The acceptance correction is evaluated using MC. The contribution to the $K_{e2}$ acceptance due to the radiative $K^+\rightarrow e^+\nu\gamma$ inner bremsstrahlung
%process was taken into account following \cite{Bijnens:1992en,Weinberg:1965nx,Gatti:2005kw}. The external bremsstrahlung suffered by the positrons in the material upstream 
%of the spectrometer magnet induced about 6\% loss of $K_{e2}$ acceptance, as a consequence of the $M^2_{miss}$ cut. This effect related to the thickness of the detector
%material has been measured using a samples of $K_{e3}$ collected simultaneusly with the $K_{e2}$ events.

The $R_K$ value is extracted from a $\chi^2$ fit to the measurements in the lepton momentum bins and data taking periods. The contributions to the systematic uncertainty include 
the uncertainties on the backgrounds, the acceptance evaluation, the beam simulation, the spectrometer alignment, the particle identification and the trigger efficiency measurement.
The result is:
\begin{equation}
  R_K\,=\,(2.488\pm0.007_{stat}\pm0.007_{syst})\times10^{-5}\,=\,(2.488\pm0.010)\times10^{-5}.
\end{equation} 
%The individual results in lepton momentum bins and per data taking period are presented in figure \ref{fig:result}. 
The result is consistent with the SM expectation and the achieved precision dominates the world average.
%\begin{figure}[htb]
%\centering
%\includegraphics[width=0.45\textwidth]{rk-vs-p-all.eps}
%\includegraphics[width=0.3\textwidth]{rk-vs-c-all.eps}
%\caption{}
%\label{fig:results}
%\end{figure}

\section{The measurement of the BR($K\rightarrow\pi\nu\nu$): future prospects at NA62}
\label{sec:pvv}
The goal for the future of the NA62 experiment is the measurement of the branching ratio of the $K^+\rightarrow\pi^+\nu\bar{\nu}$ decay with 10\%
precision. Therefore NA62 aims to collect of the order of 100 $K^+\rightarrow\pi^+\nu\bar{\nu}$ events in about two years of data taking and to 
keep the total systematic uncertainty smaller than the statistical one. To this purpose, at least $10^{13}$ $K^+$ decays are required, assuming a 10\% signal acceptance and a
$K^+\rightarrow\pi^+\nu\bar{\nu}$ branching ratio of $10^{-10}$. The need of keeping the systematic uncertainty small requires a rejection factor for generic
kaon decays of the order of $10^{12}$ and the possibility to measure efficiencies and background suppression factors directly from data.
In order to match the above required kaon intensity, signal acceptance and background suppression, new detectors must replace the existing NA62 
apparatus.
 
The same CERN-SPS extraction line already used for the $R_K$ measurement, can deliver the required intensity asking for 30\% more SPS protons on target only.
Considerations about signal acceptance drive the choice of a 75 GeV/c charged kaon beam with 1\% momentum bite. The use of a decay-in-flight technique to identify 
$K^+$ decay products is the experimental principle of NA62. 

The experimental set-up is close to the one used for the measurement of $R_K$: a 100 m beam line selects the appropriate beam; a 80 m evacuated decay volume follows 
the beam line and the detectors are placed downstream to measure the secondary particles from the kaon decays occurring in the decay volume. The secondary beam is composed of 
kaons (6\%), $\pi^+$ and protons and the rate is about 800 MHz integrated over a 16 $cm^2$ area. The rate seen by the detector downstream is about 10 MHz, mainly due to $K^+$ decays. 
The main subdetectors forming the NA62 layout are: a differential Cerenkov counter on the beam line to identify the $K^+$ in the beam; a Si-pixel beam tracker; a guard-ring counter 
surrounding the beam tracker to veto inelastic interactions of particles; a downstream spectrometer made of straw chambers in vacuum; a RICH to distinguish pions from muons; a charged 
hodoscope; a system of photon vetoes including a series of annular lead glass calorimeters surrounding the decay and detector volume, the NA48 LKr calorimeter and a small angle calorimeter 
to achieve the hermetic coverage for photons emitted at zero angle; an hadronic calorimeter to identify pions and veto muons.

\begin{figure}[htb]
\centering
\includegraphics[width=0.48\textwidth]{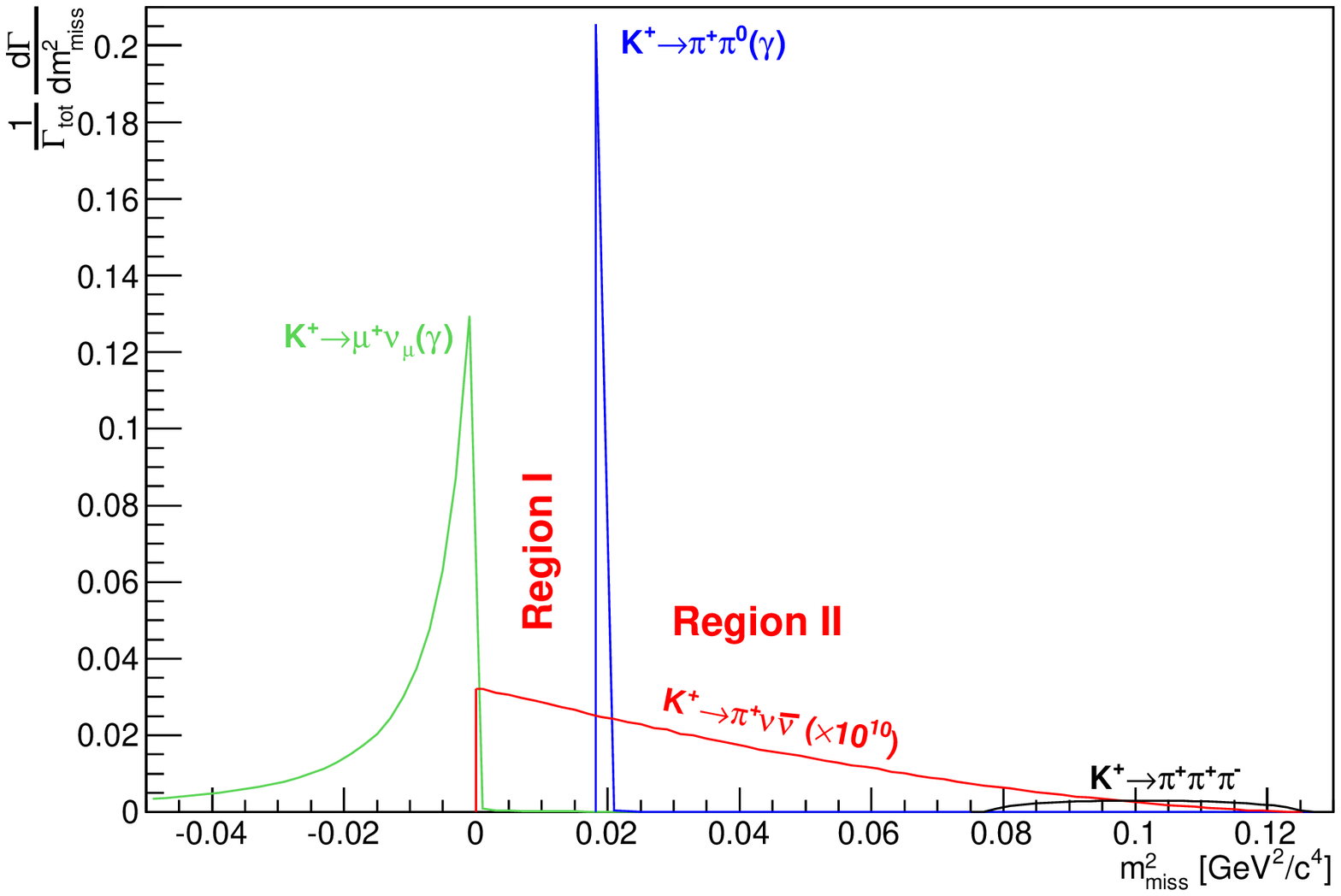}
\includegraphics[width=0.48\textwidth]{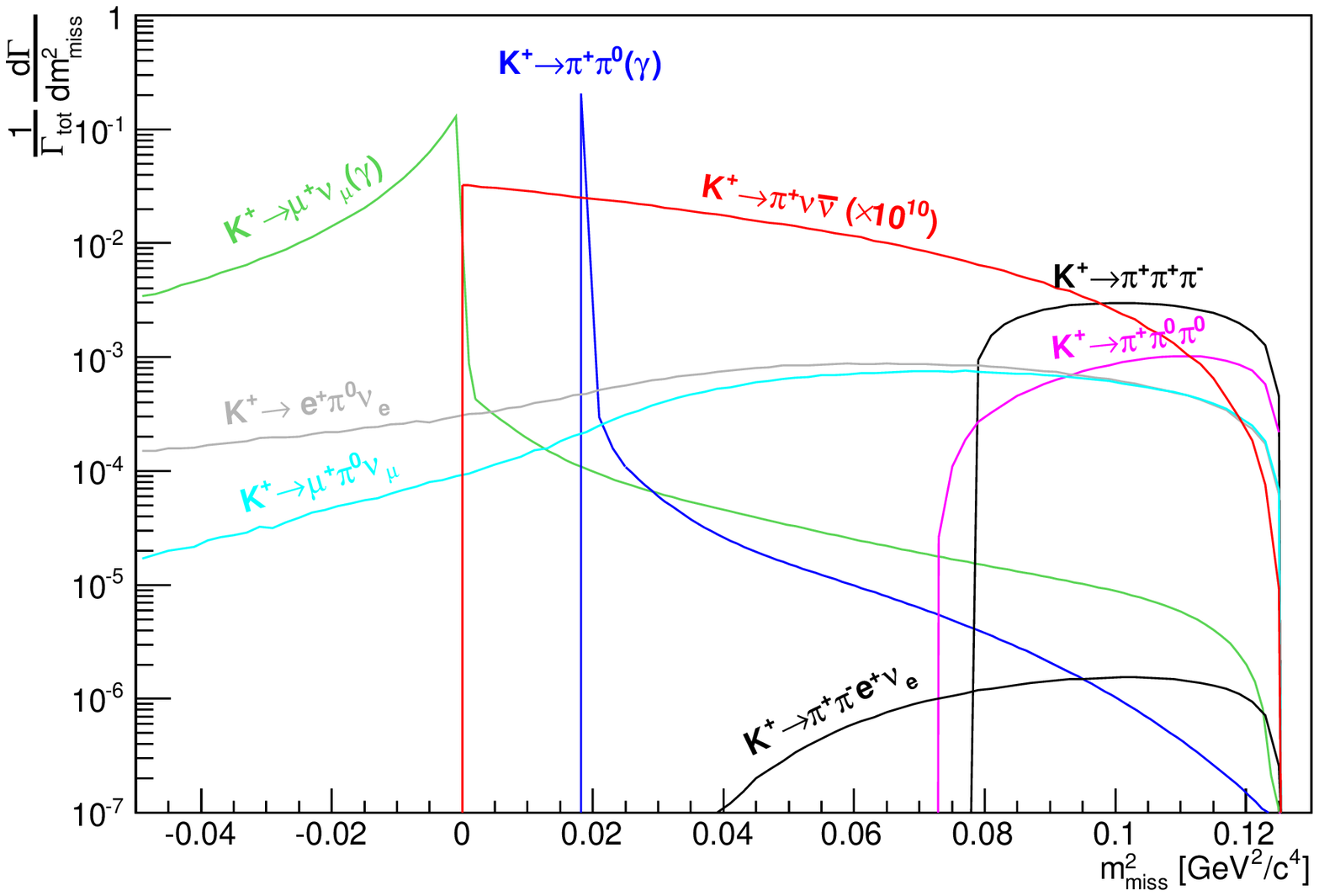}
\caption{Shape of the squared missing mass for signal and background events under the hypothesis that the charged track is a pion.}
\label{fig:pnn}
\end{figure}

The signature of the signal is one track in the final state matched to one $K^+$ track in the beam. Backgrounds come from all the kaon decays with one track left in the final state and from 
accidental tracks reconstructed downstream matched by chance to a track upstream. Let $P_K$ and $P_\pi$ be the measured four-momenta of the $K^+$ and $\pi^+$, respectively.
The variable $M^2_{miss}\equiv(P_K-P_{\pi})^2$ defines two regions (namely region I and region II) almost free of the backgrounds coming from the main $K^+$ decay modes (see fig \ref{fig:pnn}). 
The residual sources of background contamination are: tails to the $M^2_{miss}$ of the main $K^+$ decay channels due to the detector resolution and to the corresponding radiative processes; semileptonic
$K^+$ decays like $K^+\rightarrow e^+\pi^0\nu_e$; rare decays like $K^+\rightarrow\pi^+\pi^-e^+\nu_e$.

Different techniques have to be employed in combination in order to reach the required level of background rejection. Schematically they can be divided into: kinematic rejection, precise timing, high 
efficient photon vetoes, precise particle identification systems to distinguish $\pi^+$, $\mu^+$ and positrons. The kinematic rejection requires a minimal material budget for the 
kaon and pion tracking systems. The precise timing is required for a proper kaon-pion matching, given the different rate environments upstream and downstream. The high energy of the $K^+$ beam, 
the offline requirement of having a reconstructed track of maximum $35$ GeV/c momentum, a photon detector coverage up to 50 mrad and a photon detection inefficiency at the level of $10^{-5}$ for photons 
above 10 GeV in the LKr calorimeter guarantee a large enough $\pi^0$ suppression. The combined use of a RICH and a calorimetric technique allows a muon/pion separation suitable for the $K^+\rightarrow\mu^+\nu$ rejection.

Part of the subdetectors and the final beam line have been commissioned during a technical run at the end of 2012. The overall construction is on schedule for the first physics run in autumn 2014.

\section{Conclusions}
The NA62 experiment at CERN provided in 2007 the most precise measurement of the lepton flavour parameter $R_K$: $R_K=(2.488\pm0.010)\times10^{-5}$.
The ultra-rare $K\rightarrow\pi\nu\bar{\nu}$ decay is a unique environment where to search for new physics. The NA62 experiment at CERN-SPS 
will follow this road by collecting $O(100)$ events of the $K^+\rightarrow\pi^+\nu\bar{\nu}$ decay. The first physics run is scheduled for autumn 2014.


\begin{thebibliography}{99}
\bibitem{Cirigliano:2007xi}
  V.~Cirigliano and I.~Rosell,
  Phys.\ Rev.\ Lett.\  {\bf 99} (2007) 231801.

\bibitem{Masiero:2005wr}
  A.~Masiero, P.~Paradisi and R.~Petronzio,
  Phys.\ Rev.\  D {\bf 74} (2006) 011701.

\bibitem{Masiero:2008cb}
  A.~Masiero, P.~Paradisi and R.~Petronzio,
  JHEP {\bf 0811} (2008) 042.

\bibitem{Recent}
  J. Girrbach and U. Nierste, arXiv:1202.4906 (2012) ;

\bibitem{pdg}
  J. Beringer et al. (Particle Data Group), Phys. Rev. D86, 010001 (2012); 

\bibitem{Lazzeroni:2011}
  C.~Lazzeroni {\it et al.} [NA62 Collaboration],
  Phys.\ Lett.\  B {\bf 698} (2011) 105.

\bibitem{Fanti}
  V.~Fanti {\it et al.}  [NA48 Collaboration],
  Nucl.\ Instrum.\ Meth.\  A {\bf 574} (2007) 433.

\bibitem{Brod:2010hi}
  J.~Brod, M.~Gorbahn and E.~Stamou,
  Phys.\ Rev.\  D {\bf 83} (2011) 034030
  [arXiv:1009.0947 [hep-ph]].

\bibitem{Isidori:2006qy}
  G.~Isidori, F.~Mescia, P.~Paradisi, C.~Smith and S.~Trine,
  JHEP {\bf 0608} (2006) 064
  [arXiv:hep-ph/0604074].

\bibitem{Blanke:2009am}
  M.~Blanke, A.~J.~Buras, B.~Duling, S.~Recksiegel and C.~Tarantino,
  Acta Phys.\ Polon.\  B {\bf 41} (2010) 657
  [arXiv:0906.5454 [hep-ph]].

\bibitem{Artamonov:2009sz}
  A.~V.~Artamonov {\it et al.}  [BNL-E949 Collaboration],
  Phys.\ Rev.\  D {\bf 79} (2009) 092004
  [arXiv:0903.0030 [hep-ex]].

\bibitem{na62tdr}
  NA62 Technical Design Document, NA62-10-07;\\ 
  https://cdsweb.cern.ch/record/14049857.

\bibitem{Batley:2007yfa}
  J.~R.~Batley {\it et al.}  [NA48/2 Collaboration],
  Eur.\ Phys.\ J.\  C {\bf 52} (2007) 875
  [arXiv:0707.0697 [hep-ex]].

\end{thebibliography}
\end{document}